\documentstyle[12pt]{article}

\def\Journal#1#2#3#4{{#1} {\bf #2}, #3 (#4)}

\begin{document}

\title{The relativistic Calogero model\\
in an external field\thanks{Submitted to the Proceedings of the
4th Wigner Symposium, August 7-11, 1995, Guadalajara, M\'exico.}}

\author{J.F. van Diejen}

\date{}

\maketitle

\protect\vspace{-6ex}
\begin{center}
Department of Mathematical Sciences, University of Tokyo,\\
Komaba 3-8-1, Meguro-ku, Tokyo 153, Japan
\end{center}

\vspace{1ex}
\begin{abstract}
Recent results are surveyed regarding the spectrum and eigenfunctions of
the inverse square Calogero model with harmonic confinement
and its relativistic analogue.
\end{abstract}

\newpage
\section{Introduction}\label{sec1}
The Calogero model is a dynamical system that consists
of $N$ particles on the
line interacting pairwise through an inverse square potential and
coupled to a harmonic external field. Both at the level of classical and
quantum mechanics the dynamics of the system
has been studied in considerable detail
in the literature \cite{per1,op}.
Key property of the Calogero model is that,
despite the nontrivial interaction between the particles,
the equations of motion describing the
dynamics can be solved in closed form.
The exact solubility of the system stems
from the fact that it is integrable,
i.e., that there are as many
independent integrals of motion (in involution) as
degrees of freedom (viz. $N$).
For the classical model the equations of motion were
integrated by Olshanetsky and Perelomov
using a Lax pair representation, whereas
for the quantum system the spectrum and the
structure of the eigenfunctions
had already been determined before by Calogero \cite{cal}.

More recently, Ruijsenaars and Schneider introduced a relativistic
generalization of the classical
Calogero model without harmonic external field
and solved the corresponding equations of motion \cite{rs,rui1}. It was
furthermore shown that also the
relativistic system is integrable and that this
integrability is preserved after
quantization \cite{rui2}. At the quantum level,
the Hamiltonian of the relativistic model is given by a difference operator
rather than a differential operator; the nonrelativistic limit then corresponds
to sending the step size of this difference operator to zero.

Very recently the author has introduced a similar integrable relativistic
analogue of the (quantum) Calogero model with harmonic external
field \cite{die1}, and computed the corresponding spectrum and eigenfunctions
also for this case \cite{die2}.
The present contribution intends to provide an overview of these results as
well as to describe some new developments that have led to a more explicit
construction for the eigenfunctions of the nonrelativistic model \cite{bhv}.

\section{The Quantum Calogero Model}\label{sec2}
The quantum Calogero model is characterized by a Hamiltonian of the form
\begin{equation}\label{HamC}
H_{C}=  \sum_{1\leq j\leq N} \Bigl( -\frac{\partial^2}{\partial x_j^2}
+\omega^2 x_j^2 \Bigr)\;\;\; +
\sum_{1\leq j \neq k\leq N} \frac{g(g-1)}{(x_j-x_k)^2} .
\end{equation}
It is immediate from the Hamiltonian that the model may be viewed as a
system of $N$ coupled harmonic oscillators.
Should the inverse square coupling term be absent ($g=0$), then it is of course
very easy to solve the corresponding eigenvalue problem. The (boson)
eigenfunctions are in that case the product of a Gaussian ground state wave
function and symmetrized products of Hermite polynomials, and the spectrum is
that of an $N$-dimensional harmonic oscillator.
The remarkable observation by Calogero, to date around 25 years ago,
is that much of this picture is preserved after switching on the
inverse square interaction ($g> 0$) \cite{cal}.

Specifically, for the interacting system the ground state wave function becomes
\begin{equation}
\Psi_0 (\vec{ x}) =
\prod_{1\leq j< k\leq N} |x_j-x_k|^{\mbox{}^{g}}\;
\exp (-\frac{\omega}{2} \sum_{1\leq j\leq N} x_j^2 )
\end{equation}
and the ground state energy reads
\begin{equation}\label{grounde}
E_0 = \omega N (1+ (N-1)g ) .
\end{equation}
The wave functions of the excited states are again products
of the ground state wave function and certain symmetric polynomials:
\begin{equation}
\Psi_{\vec{n}} (\vec{x})=
\Psi_0 (\vec{x})\: P_{\vec{n}} (\vec{x}) ,
\end{equation}
where $\vec{n}=(n_1,\ldots ,n_N)$ denotes a vector of
(integer) quantum numbers with
$n_1\geq n_2 \geq \cdots \geq n_N\geq 0$ labeling the eigenfunctions.
The corresponding eigenvalues are given by
\begin{equation}\label{spec}
E_{\vec{n}} =E_0 + 2\omega \sum_{1\leq j\leq N} n_j .
\end{equation}

It is clear from Eq.~\ref{spec} that, apart from an overall shift in the
energy, the spectrum of the model with inverse square interaction
coincides with that of a system of {\em independent} harmonic oscillators.
Crucial difference with the latter system, however, is that
for the interacting system
the wave functions $\Psi_{\vec{n}}(\vec{x})$ do not separate in
one-particle wave functions. In fact, until recently not much
information had been available regarding the precise nature of the polynomials
$P_{\vec{n}}(\vec{x})$, except for small particle number.
(For $N\leq 5$ the polynomials $P_{\vec{n}}(\vec{x})$ were constructed
explicitly by Perelomov and Gambardella \cite{per2,gam}.)

\section{Creation and Annihilation Operators}\label{sec3}
It is well known that the wave functions for a system of independent harmonic
oscillators can be constructed by means of creation and annihilation operators.
Interestingly enough, it was recently discovered that the same classical
technique may also be applied to the case with an inverse square coupling
between the oscillators \cite{bhv}.

To this end it is convenient to introduce the concept of the so-called
Dunkl derivative \cite{dun}:
\begin{equation}
D_j =\frac{\partial}{\partial x_j}
+ g \sum_{1\leq k\leq N,\;k\neq j} (x_k-x_j)^{-1} S_{j,k} ,
\end{equation}
where $S_{j,k}$ denotes the transposition operator interchanging the particles
$j$, $k$
\begin{equation}
S_{j,k} \Psi (x_1,\ldots ,x_j,\ldots ,x_k,\ldots ,x_N)=
\Psi (x_1,\ldots ,x_k,\ldots ,x_j,\ldots ,x_N).
\end{equation}
The Dunkl derivative $D_j$ is a deformation of the ordinary derivative
$\partial_j=\partial /\partial x_j$ with the coupling constant $g$
acting as a deformation parameter.
The essential point is now that it is possible to
incorporate the inverse square coupling
between the oscillators in
the standard construction of the eigenfunctions
by means of creation and annihilation operators
if one replaces the ordinary derivative by the Dunkl derivative.
More specifically, after setting
\begin{equation}
A^\pm_j = \frac{1}{\sqrt{2}}(\mp D_j +\omega x_j) ,
\end{equation}
one has
\begin{equation}
\tilde{H}_{C} \equiv  \!\sum_{1\leq j\leq N} ( A^+_jA^-_j +A^-_jA^+_j )
=\!\sum_{1\leq j\leq N} \Bigl( -\frac{\partial^2}{\partial x_j^2}
+\omega^2 x_j^2 \Bigr)\; +\!\!
\sum_{1\leq j \neq k\leq N} \frac{g(g-S_{j,k})}{(x_j-x_k)^2}
\end{equation}
and
\begin{equation}
\left[ \tilde{H}_{C},A^\pm_j\right] =\pm 2\omega A^\pm_j.
\end{equation}
When restricted to the boson sector
(i.e., the space of permutation invariant wave functions),
the operator $\tilde{H}_C$ coincides with the Calogero Hamiltonian
$H_C$ (cf. Eq.~\ref{HamC}).
The ground state wave function in this sector is determined by its
permutation-invariance and the fact that it is annihilated by the
lowering operators
\begin{equation}
A^-_j\Psi_{0} (\vec{x})= 0 , \;\;\;\;\;\;\;\;\;\;\;\;\;\; j=1,\ldots ,N.
\end{equation}
The excited states are obtained by acting with the creation operators
on the ground state
\begin{equation}
\Psi_{\vec{n}} (\vec{x})=
\sum_{\sigma \in S^N} \Bigl( \prod_{1\leq j\leq N}
(A^+_{\sigma (j)} )^{n_j}\Bigr) \Psi_{0} (\vec{x})
\end{equation}
(where we have, in order to obtain boson
eigenfunctions, symmetrized over all permutations).
The shift of the spectrum as compared to the model without
inverse square coupling between oscillators originates
from a small change in the usual commutation relations
satisfied by the modified creation and annihilation operators $A_j^\pm$:
\begin{equation}
[A^\pm_j,A^\pm_k]=0,\;\;\;\;\;\;\;\;\;\;\;
[A^-_j,A^+_k]= \omega (1+g\sum_{1\leq l\leq N} S_{j, l} )\delta_{j,k}
-\omega g S_{j,k} .
\end{equation}
(So $H_C \Psi_0=\sum_j ( [A^-_j,A^+_j]+2A^+_jA^-_j)\Psi_0=
\sum_j \omega (1+(N-1)g)\Psi_0= E_0\Psi_0$, with
$E_0$ given by Eq.~\ref{grounde}.)

\section{Relativistic Analogue}\label{sec4}
A few years ago, Ruijsenaars introduced
an integrable quantum $N$-particle system characterized
by a (rather unorthodox)
Hamiltonian given by the second order difference operator \cite{rui2}
\begin{equation}\label{R}
H_R = \sum_{1\leq j\leq N} \left( V_j^{1/2}
\; e^{\frac{\partial}{i\partial x_j} }\;
\overline{V}_j^{1/2} \; +\;
\overline{V}_j^{1/2}
\; e^{- \frac{\partial}{i\partial x_j} }\;
 V_j^{1/2} \right) ,
\end{equation}
where $\exp (\pm \frac{\partial}{i\partial x_j} )\Psi (x_1,\ldots ,x_N)
=\Psi (x_1,\ldots ,x_{j-1},x_j\mp i,x_{j+1},\ldots x_N)$ and
\begin{equation}
V_j = \prod_{1\leq k\leq N,\; k\neq j} v(x_j-x_k),\;\;\;\;\;\;\;\;\;\;
v(z)=1+g/(iz)
\end{equation}
(with $\overline{V}_j$ denoting the complex conjugate of $V_j$).
As it turns out, Ruijsenaars' difference model may be interpreted as a
system composed of $N$ relativistic particles in $(1+1)D$
that interact with each other by means of the coefficients
$V_j$.  For $g=0$ the particles are independent ($V_j=1$), whereas
for $g>0$ each particle feels the presence of the remaining $N-1$ particles
as a change of its dynamical mass.
A more detailed discussion of the model (both at the classical and quantum
level),
with an emphasis on matters involving integrability and Poincar\'e invariance,
can be found in Ruijsenaars' papers \cite{rs,rui1,rui2}.

Quite recently, the present author observed that it is possible to introduce an
external field coupling to the relativistic system without destroying its
integrability \cite{die1}.
The Hamiltonian of the corresponding difference model reads
\begin{equation}\label{DC}
H = \sum_{1\leq j\leq N} \left( V_j^{1/2}
\; e^{\frac{\partial}{i\partial x_j} }\;
\overline{V}_j^{1/2} \; +\;
\overline{V}_j^{1/2}
\; e^{- \frac{\partial}{i\partial x_j} }\;
 V_j^{1/2} -V_j-\overline{V}_j\right) ,
\end{equation}
with
\begin{eqnarray}
V_j &=& w(x_j) \prod_{1\leq k\leq N,\; k\neq j} v(x_j-x_k),\label{V} \\
v(z)&=& 1+g/(iz),\;\;\;\;\;\; w(z)\; =\; (a+iz)\, (b+iz) \label{vw}.
\end{eqnarray}
The function $w$ encodes the external field.
For $w=1$ (this is achieved by sending the parameters $a$ and $b$
to infinity after having rescaled the Hamiltonian by
division by $ab$), the Hamiltonian $H$ reduces,
up to an irrelevant additive constant,
to the Ruijsenaars Hamiltonian $H_R$. The constant is caused by
the part $-\sum_{1\leq j\leq N} (V_j+\overline{V}_j)$,
which does not depend on $x_j$ if $w=1$ (as is readily seen
with the aid of Liouville's theorem after having inferred that
the expression is regular in $x_j$ and bounded for $x_j\rightarrow \infty$).

Below, we will discuss the spectrum and eigenfunctions of
the difference model with external field given by the Hamiltonian
in Eqs. \ref{DC}-\ref{vw} and describe its relation to
the Calogero model discussed in Sections~\ref{sec2} and \ref{sec3}.
Throughout it will be assumed that the coupling constant $g$ be nonnegative and
that (the real parts of) the parameters $a$ and $b$ be positive.

\section{Spectrum and Eigenfunctions}\label{sec5}
The ground state wave function for the difference Hamiltonian $H$ is given by
\begin{equation}\label{gstD}
\Psi_0(\vec{x}) =
\prod_{1\leq j< k\leq N} \left|
\frac{\Gamma (g +i(x_j-x_k))}{\Gamma (i(x_j-x_k))} \right|
\prod_{1\leq j\leq N} | \Gamma (a+ix_j)\, \Gamma (b+ix_j) | .
\end{equation}
Probably the simplest way to see that
this is indeed an eigenfunction is to check that conjugation
of the Hamiltonian with $\Psi_0$ yields
(using the standard difference equation $\Gamma (z+1)=z\,\Gamma(z)$ for the
gamma function) \begin{equation}
{\cal H}=\Psi_0^{-1} H \Psi_0 =
\sum_{1\leq j\leq N}
 \left( V_j
\; (e^{\frac{\partial}{i\partial x_j} }-1) \; +\;
\overline{V}_j
\; (e^{- \frac{\partial}{i\partial x_j} }-1) \right) .
\end{equation}
The transformed operator ${\cal H}$ clearly annihilates constant functions,
so $\Psi_0$ is an eigenfunction of $H$ with eigenvalue zero.

Just as for the nonrelativistic Calogero model, the wave functions
corresponding to the excited states are a product of the ground state wave
function and symmetric polynomials. More precisely, one has
\begin{equation}
H \Psi_{\vec{n}}= E_{\vec{n}}\, \Psi_{\vec{n}}
\end{equation}
with
\begin{equation}
E_{\vec{n}} =
\sum_{1\leq j\leq N} n_j (n_j + a+\overline{a}+b +\overline{b} +2(N-j)g)
\end{equation}
and
\begin{equation}
\Psi_{\vec{n}}(\vec{x}) = \Psi_0(\vec{x})\, P_{\vec{n}}(\vec{x}) ,
\end{equation}
where $P_{\vec{n}}(\vec{x})$ is the symmetric polynomial
determined by the conditions:

{\em i.} $\displaystyle P_{\vec{n}}(\vec{x}) = m_{\vec{n}}(\vec{x})
  + \sum_{\vec{n}^\prime < \vec{n}} c_{\vec{n},\vec{n}^\prime}\:
               m_{\vec{n}^\prime}(\vec{x})$;

$ii.$\
$\displaystyle
\int_{-\infty}^\infty\cdots\int_{-\infty}^\infty  P_{\vec{n}}(\vec{x})\:
\overline{m_{\vec{n}^\prime}(\vec{x})}\: \Psi_0^2 (\vec{x}) \;dx_1\cdots dx_N
=0$ \ \ if \ \ $\vec{n}^\prime < \vec{n}$.

\noindent Here the functions $m_{\vec{n}}(\vec{x})$, $\vec{n}=(n_1,\ldots
,n_N)$
with $n_1\geq n_2\geq\cdots\geq n_N\geq 0$, denote the basis consisting of
symmetrized monomials
\begin{equation}
m_{\vec{n}}(\vec{x}) = \sum_{\vec{n}^\prime \in S_N (\vec{n})}
x_1^{n_1^\prime}\cdots x_N^{n_N^\prime} ,
\end{equation}
which is partially ordered by the definition
\begin{equation}
\vec{n}^\prime \leq \vec{n} \;\;\;\;\; {\rm iff}\;\;\;\;\;
n_1^\prime +\cdots +n_k^\prime \leq n_1+\cdots +n_k
\;\;\; {\rm for}\;\;\; k=1,\ldots ,N
\end{equation}
($\vec{n}^\prime < \vec{n}$ iff $\vec{n}^\prime \leq \vec{n}$
and $\vec{n}^\prime \neq \vec{n}$).
Thus, the polynomial $P_{\vec{n}}(\vec{x})$ boils down to
the symmetrized monomial $m_{\vec{n}}(\vec{x})$
minus its orthogonal projection with respect to the $L^2$ inner product with
weight function $\Psi_0^2(\vec{x})$ onto the finite-dimensional subspace
spanned by the monomials $m_{\vec{n}^\prime}(\vec{x})$ with
$\vec{n}^\prime < \vec{n}$. For $N=1$ the resulting polynomials
are well-studied in the literature and known  as
continuous Hahn polynomials \cite{as,ask}.

The proof of the above statements hinges on a standard technique going back
(essentially) to Sutherland \cite{sut}.
First it is shown that the transformed operator ${\cal H}$ is triangular with
respect to the partially ordered monomial basis, i.e.
\begin{equation}\label{trian}
({\cal H} \, m_{\vec{n}}) (\vec{x}) =
\sum_{\vec{n}^\prime\leq \vec{n}} [{\cal H}]_{\vec{n},\vec{n}^\prime}\:
m_{\vec{n}^\prime}(\vec{x}) ,
\end{equation}
where the $[{\cal H}]_{\vec{n},\vec{n}^\prime}$ represent
certain (complex) matrix elements.
It is clear that acting with ${\cal H}$ on a monomial $m_{\vec{n}}$ yields a
permutation invariant rational function. The permutation symmetry guarantees
that the simple poles at $x_j=x_k$ (caused by the zero in the denominator of
$v(z)$) all cancel each other. Hence, the rational function is actually a
(permutation invariant) polynomial and can thus indeed be expanded in
symmetrized monomials. To see that in this expansion only monomials
$m_{\vec{n}^\prime}$
with $\vec{n}^\prime \leq \vec{n}$ occur (triangularity), one uses the
asymptotics at infinity.
Setting $x_j=R^{y_j}$ with $y_1>y_2>\cdots >y_N>0$
gives the following asymptotics for $R\rightarrow +\infty$:
\begin{eqnarray}\label{as1}
m_{\vec{n}^\prime}&=& R^{\vec{n}^\prime \cdot\vec{y}} \;+
o(R^{\vec{n}^\prime\cdot\vec{y}}), \\
{\cal H} m_{\vec{n}}&=& O(R^{\vec{n}\cdot\vec{y}}) .\label{as2}
\end{eqnarray}
By comparing the asymptotics of Eqs.~\ref{as1} and \ref{as2}, and using
the fact that
\begin{equation}
\vec{n}^\prime \leq \vec{n} \;\;\;\; {\rm iff}\;\;\;\;
\vec{n}^\prime \cdot \vec{y} \leq \vec{n}\cdot\vec{y}\;\;\;\;\;\;\;
\forall \vec{y}\;\;\; {\rm with}\;\;\; y_1>y_2>\cdots >y_N>0,
\end{equation}
one infers that the matrix elements $[{\cal H}]_{\vec{n},\vec{n}^\prime}$
in the expansion of ${\cal H} m_{\vec{n}}$ in terms of
symmetrized monomials $m_{\vec{n}^\prime}$
can only be nonzero if $\vec{n}^\prime \leq \vec{n}$,
which gives Eq.~\ref{trian}.

Next one observes that since $H$ is Hermitian the transformed operator
${\cal H}$ is symmetric with respect to the $L^2$ inner product with weight
function $\Psi_0^2(\vec{x})$:
\begin{eqnarray}
\lefteqn{ \int_{-\infty}^\infty\cdots\int_{-\infty}^\infty
({\cal H}m_{\vec{n}})(\vec{x})\:
\overline{m_{\vec{n}^\prime}(\vec{x})}\: \Psi_0^2 (\vec{x}) \;dx_1\cdots dx_N
=} & & \nonumber \\
& &\int_{-\infty}^\infty\cdots\int_{-\infty}^\infty
m_{\vec{n}}(\vec{x})\:
\overline{({\cal H}m_{\vec{n}^\prime})(\vec{x})}\: \Psi_0^2 (\vec{x})
\;dx_1\cdots dx_N .
\end{eqnarray}

By combining the triangularity and symmetry of ${\cal H}$ it is not difficult
to deduce that
${\cal H}P_{\vec{n}}$ is a linear combination of monomials $m_{\vec{n}^\prime}$
with $\vec{n}^\prime \leq \vec{n}$, which is orthogonal to
$m_{\vec{n}^\prime}$ with $\vec{n}^\prime < \vec{n}$. Hence,
${\cal H}P_{\vec{n}}$ must be proportional to $P_{\vec{n}}$, i.e.,
$P_{\vec{n}}$ is an eigenfunction of ${\cal H}$. The corresponding eigenvalue
$E_{\vec{n}}$ is obtained by computing the diagonal matrix element
$[{\cal H}]_{\vec{n},\vec{n}}$, i.e., by explicitly computing the leading
coefficient in
Expansion~\ref{trian}.

\section{The Nonrelativistic Limit}\label{sec6}
To relate the difference model to the Calogero model one needs to explicitly
introduce a step size parameter by rescaling the positions $x_j$ and the
parameters $a$, $b$. Substituting
$x_j\rightarrow \beta^{-1} x_j$ (so
$\partial_j\rightarrow \beta\partial_j$),\
$a\rightarrow (\beta^2\omega)^{-1}$, \
$b\rightarrow (\beta^2\omega^\prime )^{-1}$, and multiplying
$H$ by $\beta^2\omega\omega^\prime$, leads to
a Hamiltonian $H$ given by Eqs.~\ref{DC}, \ref{V} with
$\exp (\pm i\partial_j)$ replaced by $\exp (\pm i\beta \partial_j)$ and
functions $v$, $w$ of the form
\begin{equation}
v(z) = 1 +\beta g/(iz),\;\;\;\;\;\;\;\;\;\;\;\;\;
w(z) = \beta^{-2} (1+ i\beta\omega\, z)
                  (1+ i\beta\omega^\prime \, z) .
\end{equation}
The step size parameter $\beta$ should be compared with the inverse of the
light speed appearing in Ruijsenaars' model \cite{rui2}.

For $\beta\rightarrow 0$ one now has
\begin{equation}
H \rightarrow
\sum_{1\leq j\leq N} \Bigl( -\frac{\partial^2}{\partial x_j^2}
+(\omega+\omega^\prime)^2 x_j^2 \Bigr)\;\;\; +
\sum_{1\leq j \neq k\leq N} \frac{g(g-1)}{(x_j-x_k)^2} - \varepsilon_0,
\end{equation}
with $\varepsilon_0=(\omega+\omega^\prime)N(1+(N-1)g)$, and
$E_{\vec{n}} \rightarrow 2(\omega+\omega^\prime)\sum_{1\leq j\leq N} n_j$.
The wave functions go (after dividing by a divergent numerical factor
arising from the gamma factors in $\Psi_0$, Eq.~\ref{gstD}) over in
$\Psi_{\vec{n}}(\vec{x})=\Psi_0(\vec{x}) P_{\vec{n}}(\vec{x})$, where
\begin{equation}\label{gst}
\Psi_0(\vec{x}) =\prod_{1\leq j< k\leq N} |x_j-x_k|^g\,
    \prod_{1\leq j\leq N} e^{-\frac{1}{2}(\omega+\omega^\prime)\, x_j^2 }
\end{equation}
and $P_{\vec{n}}(\vec{x})$ is the polynomial determined by the Conditions
{\em i.} and {\em ii.} of Section~\ref{sec5} with $\Psi_0$ now taken from
Eq.~\ref{gst}.
This way we recover for $\beta\rightarrow 0$ the Hamiltonian, the spectrum and
the eigenfunctions of the Calogero model.

\vfill
\section*{Acknowledgments}
This work was made possible by financial support from
the Japan Society for the Promotion of Science (JSPS).

\end{document}